\documentclass{appolb}

\usepackage{hyperref}

\usepackage{amsmath}
\usepackage{slashed}
\usepackage{graphicx}

\newcommand{\beq}{\begin{equation}}
\newcommand{\eeq}{\end{equation}}
\newcommand{\bea}{\begin{eqnarray}}
\newcommand{\eea}{\end{eqnarray}}

\newcommand{\xu}{x_{1}}
\newcommand{\xd}{x_{2}}
\newcommand{\xn}{x_{n}}

\newcommand{\pd}{\partial}

\begin{document} 

\title{Vacuum correlations of the stress-energy-momentum tensor with constituent quarks} 

\author{Mirko Serino$^1$\thanks{mirkos.serino@gmail.com}, 
            Wojciech Broniowski$^{2}$\thanks{wojciech.broniowski@ifj.edu.pl}, 
            Enrique Ruiz Arriola$^{3}$\thanks{earriola@ugr.es} 
\address{$^{1}$Lungomare Matteotti 23, 73026, Torre dell’Orso, Melendugno (LE), Italy}
\address{$^{2}$The H. Niewodnicza\'nski Institute of Nuclear Physics, Polish Academy \\ of Sciences, 31-342~Cracow, Poland}
\address{$^{3}$Departamento de F\'{\i}sica At\'{o}mica, Molecular y Nuclear and Instituto Carlos I \\ de  F{\'\i}sica Te\'orica y Computacional}
}

\maketitle

\begin{abstract}
The two point correlation function of the stress-energy-momentum
tensor describes the propagation of a space-time ``micro-earthquake''
in the vacuum.  In the framework of the path integral formulation of
field theory in curved space-time, we discuss the Ward-Takashi
identity for two-point Green's function of the stress-energy-momentum
tensor for a general case of a non-conformal theory.  The identity
constrains the longitudinal part of the correlator, with the vacuum
expectation value of the stress-energy-momentum, non-zero in a
non-conformal theory.  The obtained formula is demonstrated on the
free massive Dirac fermion theory, treated at the one-loop level. This
example befits a class of phenomenological chiral quarks models which
have been used successfully in numerous applications in the soft
non-perturbative regime of strong interactions. We stress the role of
the constraints following from the Ward-Takahashi identity for the
correlation functions in these models and point out the differences
with the low-energy theorems (sum rules), not complying to the
symmetry requirements.  We also show the temporal representation
of the two-point correlators, which is an object amenable to lattice
QCD simulations.
\end{abstract}

\date{29 September 2026}

\section{Introduction \label{sec:intro}}

The stress-energy-momentum tensor (SEM) is a conserved quantity in
relativistic quantum field theory and, consequently, has a long
history. The canonical Noether construction has been amended by
Belinfante and Rosenfeld~\cite{Belinfante:1940,rosenfeld1940tenseur},
allowing to obtain a symmetric form while respecting the conservation
laws.  With torsion-free connections, this prescription is equivalent
to the Hilbert construction, where SEM is the response of the action
to perturbations of the metric tensor in curved space-time about its
flat-space value (for a recent discussion of these issues see, e.g,
\cite{Blaschke:2016ohs,Baker:2020eqs}).  Further, within this
framework, multiple vacuum SEM correlations (Green's functions) arise
as multilinear responses to these perturbations and, in that regard,
they provide basic information on the gravitational space-time
fluctuations (see, e.g.,~\cite{Birrell:1982ix}). For the rest of the
paper we will embrace the physically motivated Hilbert interpretation
of SEM as the response of the system to the presence of a space-time
fluctuation in the metric caused either by gravity or other external
sources.

The theoretical and practical phenomenological implications involving
the SEM are vast, so we briefly mention only some of them to provide a
motivating framework for our study.  Since the early days of the deep
inelastic scattering, the momentum sum rule, corresponding to the
expectation value of SEM in a hadron, has played a major role, as it
provides a measure of the relevance of the valence quarks, sea quarks,
and gluons in a hadron state (see e.g.  \cite{pokorski2000gauge} for a
comprehensive discussion). This connection to the parton distribution
functions (PDFs), the generalized parton distribution functions (GPDs)
and the corresponding gravitational form factors has been explored in
lattice QCD simulations of hadron structure (for a recent 
lattice and theoretical progress on the gravitational form factors of
the pion and nucleon,
see~\cite{Hackett:2023rif,Hackett:2023nkr,RuizArriola:2026wkb} and
numerous references therein).

Two-point vacuum correlators of SEM have been studied in the context of
the operator product expansion (OPE) in the QCD sum rules for gluonium
states by Novikov, Shifman, Vainshtein, and
Zakharov~\cite{Novikov:1981xi}  in phenomenological studies of the
$J^{PC}=2^{++}$ states.  A dispersive analysis of the glueball masses was carried out within a method based on OPE in~\cite{Li:2021gsx}. 
The mixed scalar-SEM correlator was explored
within Chiral Perturbation Theory by Donoghue and
Leutwyler~\cite{Donoghue:1991qv}.

At finite temperatures, the two-point correlators of SEM are related
to viscosities via the Kubo formulas.  These correlators are  basic
objects in the lattice QCD quest to find the transport coefficients of
hot QCD medium~\cite{Meyer:2009jp,Taniguchi:2017ibr}.  The shear
viscosity of a hot medium in a chiral quark model was calculated
in~\cite{Lang:2013lla}.

On a more fundamental though less practical setup, the $n$-point SEM
Green's functions describe a multi-graviton scattering process, but
the smallness of the gravitational constant and no access to
gravitons, obviously, precludes a direct experimental information.  
Nevertheless, the case of {\em conformal} field theories has been quite thoroughly studied, starting with
Osborn and collaborators' pioneering works on the two- and three-point functions in the coordinate 
space~\cite{Osborn:1993cr,Erdmenger:1996yc}. Extension to the momentum space followed, first via the Fourier transform~\cite{Coriano:2012wp}, 
and later with an approach contained entirely in the momentum space~\cite{Bzowski:2013sza,Bzowski:2017poo}.
The case of the scalar field has been considered by one of us (MS) in the momentum space for the four-point SEM correlation function~\cite{Serino:2014ysa,Serino:2020pyu}.

Unlike other conserved quantities such as the electromagnetic or
flavor currents, SEM carries additional complications
from higher-order space-time derivatives of the fields, which requires
introducing the $T^\ast$ time-ordering~\cite{nambu1952lagrangian}.
The path-integral framework, as used in this paper, is equivalent to
choosing a certain $T^\ast$ time-ordered product.  Also, the correlations of
SEM induce a highly divergent high-energy behavior.  In a pioneering
paper, Callan, Coleman, and Jackiw~\cite{Callan:1970ze} showed how to
alter the construction in such a way that a new improved tensor not
only has a better ultraviolet behavior, but its source is the
gravitational field itself, in harmony with the equivalence principle.

We believe a more phenomenological and practical discussion in the
case of strong interactions seems worth and insightful, as it
corresponds with the fluctuating mass distribution properties of
matter and the vacuum subjected to proper symmetry constraints. In
this paper we first derive, with path integrals in curved space-time,
the Ward-Takashi identity (WTI) for two-point Green’s function of SEM
for a general case of a {\em non-conformal} field theory.  We then
analyze the case of free massive Dirac fermions. With appropriate
Feynman rules, we obtain an explicit one-quark-loop formula for the
two-point function which complies to the general symmetry
requirements, in particular WTI.  Such fermionic systems with quarks
have been used extensively to model low-energy dynamics of strong
interactions, where quarks acquire a large (constituent) mass due to
the dynamical chiral symmetry breaking induced in QCD (for a review,
see~\cite{RuizArriola:2002wr}). The model has no gluons at the
co-called quark-model scale~\cite{RuizArriola:2002wr}, denoted as $\mu_{\rm QM}\simeq 320~{\rm MeV}$, where massive
quarks are its only degrees of freedom (for early studies on the
interplay of the chiral quark models and SEM within chiral effective
Lagrangians see,
e.g.,~\cite{Andrianov:1998fr,Megias:2004uj,Megias:2005fj}).  Finally,
with possible lattice QCD simulations~\cite{CARACCIOLO1990119,Bresciani:2026fqc,Meyer:2008gt} in mind, we discuss the quark-model SEM correlators
in the temporal representation, where only their absorptive parts
contribute.

\section{Correlators and the gravitational Ward-Takahashi identities \label{sec:wti}}

Our notation follows closely Chapter~2 of~\cite{Serino:2014ysa}, adapted to the Minkowski space-time convention. The standard Hilbert definition of SEM in a classical field theory described by an action $\mathcal S$ of the matter fields is 
given in terms of a functional derivative with respect to the metric tensor $g_{\mu\nu}(x)$, once the theory has been 
embedded in a curved space-time, 
\beq \label{emt}
\theta^{\mu\nu}(x) = -\frac{2}{\sqrt{g_{x}}}\frac{\delta \mathcal{S}}{\delta g_{\mu\nu}(x)}, 
\eeq
where $g_x \equiv -\textrm{det}\, g_{\mu\nu}(x)$. The generating functional is given by the path integral
\beq
\mathcal Z=e^{i \mathcal W} = \frac{1}{\mathcal{N}} \, \int \, \mathcal D\Phi \, e^{i \mathcal S}, \label{eq:w}
\eeq
where ${\mathcal W}$ generates the connected diagrams, $\mathcal{N}$ is a normalization, 
and $\Phi$ denotes all the quantum fields of the theory.

The $n$-point connected correlators of SEM are obtained as
\begin{eqnarray}
  && \langle \theta^{\mu_1\nu_1}(\xu)\dots \theta^{\mu_n\nu_n}(\xn)\rangle = \label{eq:cdef} \\ 
&& \hspace{4mm} 
 \frac{(2i)^n}{{\sqrt{g_{\xu}}\dots \sqrt{g_{\xn}}}}\,  \frac{i \, \delta^n  \mathcal{W}}{\delta g_{\mu_1\nu_1}((\xu))\dots \delta g_{\mu_n\nu_n}((\xn)}, \nonumber
\end{eqnarray}
where the brackets  $ \langle . \rangle$ denote the path-integral expectation value in the 
curved space-time. Note the symmetry property with the indices interchange, $(\mu_i,\nu_i) \leftrightarrow (\nu_i,\mu_i)$, following from the definition. 
As already mentioned, in this formulation $ \langle . \rangle$
defines a  particular $T^\ast$ time-ordered product.
The first two connected $n$-point functions are given explicitly in terms of the action ${\mathcal S}$ as 
\beq
\langle \theta^{\mu\nu}(x) \rangle = \int \mathcal{D} \Phi \, \frac{2i}{\sqrt{g_{x}}} \frac{\delta (i \mathcal{S})}{\delta g_{\mu\nu}(x)}\, e^{i\mathcal S} 
\label{emtVev}
\eeq
and
\begin{eqnarray}
&& \hspace{-7mm} \langle \theta^{\mu\nu}(\xu)\theta^{\alpha\beta}(\xd)\rangle = \int \mathcal D \Phi\, e^{i\mathcal S} 
\frac{(2i)^2}{{\sqrt{g_{\xu}} \sqrt{g_{x_2}}}}\times \\
&& \left[\left\langle \frac{\delta (i \mathcal S)}{\delta g_{\mu\nu}(\xu)}\frac{\delta (i \mathcal S)}{\delta g_{\alpha\beta}(\xd)}\right\rangle + \left\langle \frac{\delta^2 (i \mathcal S)}{\delta g_{\mu\nu}(\xu)\delta g_{\alpha\beta}(\xd)}\right\rangle \right].  \nonumber
\end{eqnarray}
Note the presence of the tadpole piece (the second term).

We remark that in the presence of fermions, one usually uses the formulation with the Vierbein (tetrad) fields $V^\mu_A(x)$, which encode
the transformation properties of the Dirac gamma matrices and the spin connection, allowing for a more efficient treatment of their variation 
with $g^{\mu \nu}$. Then the equivalent of Eq.~(\ref{emt}) is~\cite{Birrell:1982ix}
\beq \label{emttet}
  \theta^{\mu \nu} = \frac1{2V (x)}  \left[ V^\mu_A (x) \frac{\delta S}{\delta V^\nu_A (x)}
    + V^\nu_A (x) \frac{\delta S}{\delta V^\mu_A (x)} \right],
\eeq
where symmetrization is being imposed. Using Eq.~(\ref{emttet}) yields the same structure of WTI identities as Eq.~(\ref{emt})~\cite{Coriano:2012wp}.
A detailed derivation of SEM in the Vierbein formalism can be found, e.g., in~\cite{Alcubierre:2025wgj}.

The derivation of the gravitational WTI for the two-point function follows exactly~\cite{Coriano:2012wp,Serino:2014ysa,Serino:2020pyu}.
The  general covariance in a curved space implies the master equation
\beq
\nabla_\nu \langle \theta^{\mu\nu}(\xu) \rangle = \nabla_\nu \left(\frac{2i}{\sqrt{g_{\xu}}}\frac{i \, \delta \mathcal{W}}{\delta g_{\mu\nu}(\xu)}\right) = 0, \label{eq:master}
\eeq
By expanding the covariant derivative and using the symmetry properties of the Christoffel symbols $\Gamma^\mu_{\kappa\nu}$ one gets
\beq
\partial^{x_1}_\nu \frac{\delta \mathcal W}{\delta g_{\mu\nu}(\xu)}+\Gamma^\mu_{\kappa\nu}(\xu)\frac{\delta\mathcal W}{\delta g_{\kappa\nu}(\xu)} = 0.
\eeq
Carrying out a functional derivative of this relation with respect to $\delta g_{\alpha\beta}$ and taking the flat space-time limit (where 
$\Gamma^\mu_{\kappa\nu}=0$ but ${\delta \Gamma^\mu_{\kappa\nu}}/{\delta g_{\alpha \beta}}$ is in general non-zero) yields (cf.~Eq.~(30) of \cite{Coriano:2012wp})
\beq
\nabla_\nu \langle  \theta^{\mu\nu} (x_1) \theta^{\alpha \beta} (x_2)  \rangle_c = 
 \left[ \partial^{x_1}_\nu \frac{ \delta^2\mathcal W}{\delta g_{\mu\nu}(\xu)\delta g_{\alpha\beta}(\xd)} + 
\frac{\delta \Gamma^\mu_{\kappa\nu}(\xu)}{\delta g_{\alpha \beta}(\xd)}\frac{\delta\mathcal W}{\delta g_{\kappa\nu}(\xu)}\right ]_{g=\eta} = 0.
\eeq
Next, we perform the Fourier transform to the momentum space (all the momenta are
conventionally taken to be incoming), which together with definition (\ref{eq:cdef}) yields the WTI in the form (from
now on, taking the $g=\eta$ limit is understood, with $ \eta={\rm diag}(1,-1,-1,-1)$)
\beq
q_\nu \langle \theta^{\mu\nu}(q)\theta^{\alpha \beta}(-q) \rangle + 
   2 \frac{\delta \Gamma^\mu_{\kappa\lambda}}{\delta g_{\alpha \beta}}(q) \langle \theta^{\kappa\lambda}\rangle  = 0, \label{eq:wti2}
\eeq
The derivative of the Christoffel symbol in the momentum space is
\begin{equation}
\frac{\delta \Gamma^\mu_{\kappa \nu}}{\delta g_{\alpha \beta}}(q) = 
\frac{1}{2}\, \eta^{\mu\lambda}\left(  {s^{\alpha \beta}}_{\kappa\nu}\, q_\lambda - {s^{\alpha \beta}}_{\kappa\lambda}\,  q_\nu - {s^{\alpha \beta}}_{\lambda\nu}\, q_\kappa \right), 
\end{equation}
with ${s^{\mu\nu}}_{\alpha\beta} \equiv {\delta g_{\alpha\beta}}/{\delta g_{\mu\nu}}= 
\frac{1}{2}\, \left( \delta^{\mu}_\alpha\,\delta^{\nu}_\beta + \delta^{\nu}_\alpha\,\delta^{\mu}_\beta \right)$.
By covariance, the vacuum expectation value of SEM is proportional to the flat-space metric tensor, 
\begin{eqnarray}
\langle \theta^{\mu\nu}\rangle=B\, \eta^{\mu\nu},
\end{eqnarray} 
hence Eq.~(\ref{eq:wti2}) becomes
\begin{eqnarray}
&& \hspace{-7mm} q_\nu \langle \theta^{\mu\nu}(q)\theta^{\alpha\beta}(-q) \rangle 
= - B \left ( q^\mu \eta^{\alpha \beta} - q^\alpha \eta^{\mu \beta} - q^\beta \eta^{\mu \alpha}\right ), \label{eq:wti3}
\end{eqnarray}
which is the key result. It can equivalently be written as
\begin{eqnarray}
  q_\nu \langle \theta^{\mu\nu}(q)\theta^{\alpha\beta}(-q) \rangle 
  &=& - q^\mu \langle \theta^{\alpha \beta}\rangle + q^\alpha \langle \theta^{\mu \beta}\rangle + q^\beta \langle \theta^{\mu \alpha} \rangle \nonumber \\
&=& (- q^\mu g^{\alpha \beta} + q^\alpha g^{\mu \beta} + q^\beta g^{\mu \alpha}) B. \label{eq:wti4}
\end{eqnarray}
In non-conformal theories, such as in the case of massive fermions described in the following, 
$B\neq 0$ and the WTI has a non-trivial form, involving on the right hand side the vacuum expectation values of SEM. 
We note that the tensorial structure of Eq.~(\ref{eq:wti3}) agrees with Eq.~(15) in~\cite{Boulware}, obtained via old-fashioned techniques 
with an involved analysis of the Schwinger terms, as well as with Eq.~(6) of~\cite{Brown-Zee}. 

We will later consider also the correlation of SEM with the quark condensate\footnote{In our notation, $\bar{q}q$ denotes the quark condensate of a single flavor.} 
(scalar field) in a theory with massive quarks. This 
correlator is defined as~\cite{Donoghue:1991qv}
\begin{eqnarray}
\langle \theta^{\mu\nu}(q) \, \bar{q}q(-q) \rangle = i\!\! \int \!\!d^4x \, e^{-i q x} \langle \theta^{\mu\nu}(x)\bar{q}q(0)\rangle. \label{eq:ts}
\end{eqnarray}
Introducing a scalar source to the generating function~\cite{Freedman:1974gs} and repeating the steps outlined above one obtains
the corresponding WTI in the simple form
\begin{eqnarray}
q_\mu \langle \theta^{\mu\nu}(q) \bar{q}q(-q) \rangle = -q^\nu \langle \bar{q}q \rangle. \label{eq:wtiqq}
\end{eqnarray}

\section{Identity with the trace of SEM \label{sec:trace}}

In this paper, we use the short-hand notation for the trace operator,
\begin{eqnarray}
\theta \equiv \theta^\mu_\mu.
\end{eqnarray}

In curved space, in general, one can write
\begin{eqnarray}
g_{\mu\nu}(x) \langle \theta^{\mu\nu}(x) \rangle = \mathcal{A}[g,x] + \langle \theta(x) \rangle, \label{eq:tr0}
\label{emtvevtrace}
\end{eqnarray}
where  $\mathcal{A}$ denotes the anomalous terms which account for the difference between the trace operation taken before and after quantization (for their explicit form see, e.g., \cite{Birrell:1982ix}). Similarly as in the derivation of Sec.~\ref{sec:wti}, we 
differentiate functionally Eq.~(\ref{emtvevtrace}) with respect to $\delta g_{\alpha \beta}(x')$, take the flat space-time limit, and pass to the momentum space, which yields 
\begin{eqnarray}
\eta_{\mu\nu} \langle \theta^{\mu\nu}(q) \theta^{\alpha\beta}(-q) \rangle = 
2\, i\, \frac{\delta \mathcal{A}}{\delta g_{\alpha\beta}}[q] + \langle \theta(q) \theta^{\alpha\beta}(-q)\rangle
+ 2\, \langle \theta^{\alpha\beta} \rangle. \label{eq:trid}
\end{eqnarray}
Note a non-trivial appearance of the last term on the right-hand side, originating from $\delta g_{\mu\nu}(x)/\delta g_{\alpha \beta}(x')$.

\section{Constituent quarks}

As an illustration of the fulfillment of the above formal requirements,
in particular the WTIs and  the trace identity, we consider a model with free massive
(constituent) quarks.  Such theories, notably, are interesting in
their own right due to numerous applications to the low-energy hadronic
phenomenology~(for a review see, e.g.,~\cite{RuizArriola:2002wr}), in
particular the non-perturbative structure of pions and their
interactions. The light-flavor quarks acquire a large (constituent) quark
mass $M$ due to the spontaneous chiral symmetry breaking occurring in
QCD.  At the low-energy quark-model scale the quarks are assumed to be
the only dynamical degrees of freedom, with gluons generated
radiatively upon evolution to higher
scales~\cite{Davidson:1994uv}. Obviously, the theory is not conformal
at nonzero $M$.

The action of our theory embedded in a curved space-time, 
written with the help of the Vierbein $V_a^{\mu}(x)$, is
\begin{equation}
S\!=\!\!\int \!\! d^4 x \sqrt{-g} \left\{\tfrac{i}{2}\left[\bar\psi\gamma^a\,V_a^\mu \nabla_\mu \psi - 
(\nabla_\mu\bar\psi)\gamma^a\,V_a^\mu \psi\right] \!- \!M\bar\psi\psi \right\}. 
\label{fermionAction}
\end{equation}
The Vierbein fields satisfy the relation
$V_{a}^{\mu}(x)\, V_{b}^{\nu}(x) \, g_{\mu\nu}(x) = \eta_{ab}$. 
The covariant (world) derivative acting on a fermion field is defined 
as~\cite{Birrell:1982ix} $\nabla_\mu = \partial_{\mu} + \tfrac{1}{2}\Sigma^{a b}\, {V_{a}}^{\sigma}\,\nabla_{\mu} V_{b \sigma}$,
with $\Sigma^{a b}=\frac{1}{4}[\gamma^a,\gamma^b]$, 
whereas $\nabla_{\nu} A^\mu = \pd_\nu A^\mu + \Gamma^\mu_{\nu\rho}A^\rho$ when acting on a vector field. 
Then, Eqs.~(\ref{emttet}) and (\ref{fermionAction}) yield the SEM operator
\begin{eqnarray}
\hspace{-3mm} \theta^{\mu\nu}(x) &=& \tfrac{i}{4}\left[ \bar\psi(x)\gamma^{\nu}\nabla^{\mu} \psi(x)- (\nabla^{\mu}\bar\psi(x))\gamma^{\nu} \psi(x) + (\mu \leftrightarrow \nu) \right] \label{fermionEMT} \\ 
&-& g^{\mu\nu}(x) 
\tfrac{i}{2}\left \{ \left[ \bar\psi(x) \gamma_{\mu}\nabla^{\mu} \psi(x)- (\nabla^{\mu}\bar\psi(x))\gamma_{\mu} \psi(x)\right] - M \bar\psi(x)\psi(x) \right\}. \nonumber 
\end{eqnarray}

\begin{figure}[tb]
\begin{center}
\hspace{4mm} \includegraphics[angle=0,width=0.4 \textwidth]{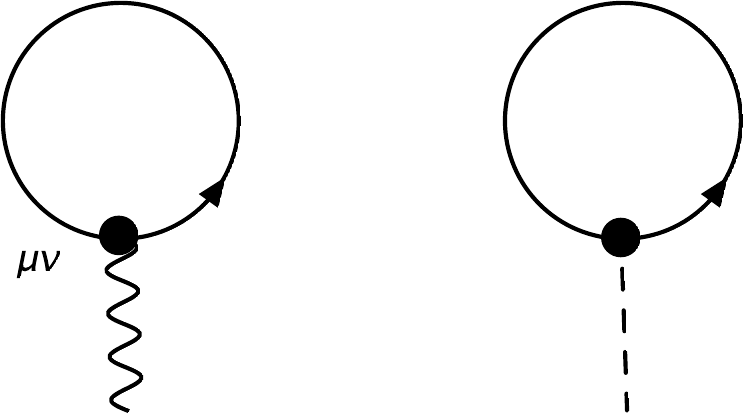} \\~\\
\includegraphics[angle=0,width=0.5 \textwidth]{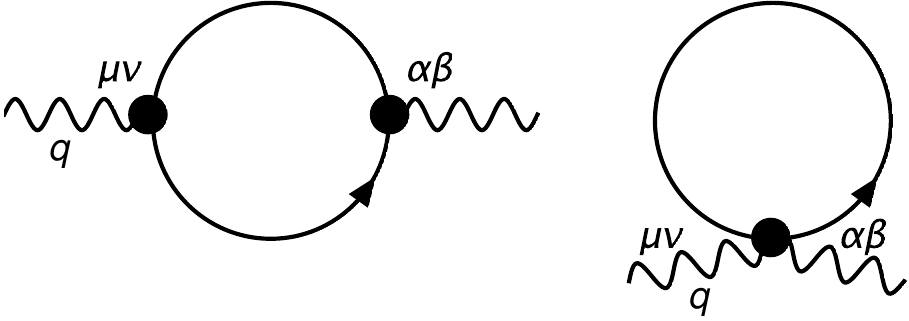} 
\end{center}
\vspace{-2mm}
\caption{One-loop Feynman diagrams for the one- and two-point functions in the constituent quark model. 
The wavy line indicates the graviton, the dashed line the scalar field, and  the solid line is the quark 
propagator with the constituent mass $M$. \label{fig:tad}} 
\end{figure}

We shall treat the model at the one-quark loop level, which corresponds to taking the large-$N_c$ limit. The Feynman diagrams for the one- and two-point 
correlators are given in Fig.~\ref{fig:tad}.
The Feynman rules for the vertices are as follows: the vertex for the incoming graviton of momentum $q$ and incoming quark of momentum $k$ is
\begin{eqnarray}
&&  \hspace{-7mm}  V_{gq}^{\mu \nu}=\tfrac{1}{4} \left [ \gamma^\mu (2k+q)^\nu - 
 \eta^{\mu \nu} (2\slashed{k}+\slashed{q}-2M) \right ] + (\mu \leftrightarrow \nu), \nonumber
\\ \label{eq:vgq}
\end{eqnarray}
whereas for the contact graviton-graviton-quark vertex we have 
\begin{eqnarray}
&& \hspace{-7mm} V_{ggq}^{\mu \nu, \alpha \beta}=
\left( \slashed{k}-M \right) \left( {\eta}^{\alpha \beta } {\eta}^{\mu \nu } 
- {\eta}^{\alpha \nu } {\eta}^{\beta \mu} -{\eta}^{\alpha \mu } {\eta}^{\beta \nu } \right) +  \label{eq:vggq} \\
&& \Big[ \left \{ \tfrac{1}{8} k^{\alpha } \left(3 \gamma^{\nu } g^{\beta \mu }+3 \gamma^{\mu } g^{\beta \nu }-4 \gamma^{\beta } g^{\mu \nu }\right) + (\alpha \leftrightarrow \beta) \right \}  +  (\mu \nu \leftrightarrow \alpha \beta) \Big] . \nonumber
\end{eqnarray}
Here, for the compactness of the notation, the symmetrization in the Lorentz 
indices has to be carried out as indicated. Also, we have taken the special case occurring in the tadpole diagrams 
in Fig.~\ref{fig:tad}, where the incoming and outgoing quark momenta are equal to $k$, which leads to simplifications. 
The intricate form of Eq.~(\ref{eq:vggq}) follows 
from the covariant structure of the action~(\ref{fermionAction}) and is crucial for the satisfaction of the WTI~(\ref{eq:wti3}). 
For the evaluation of the mixed SEM-scalar correlator we also need
\begin{eqnarray}
  V_{\sigma q}= 1 \, , \qquad  V_{g\sigma q}^{\mu \nu}= -\eta^{\mu \nu}.
\end{eqnarray}

\section{One-point functions}

The one-loop vacuum expectation value of SEM 
with massive quarks is evaluated according to the top-left diagram in Fig.~\ref{fig:tad}, yielding
\begin{eqnarray}
\langle \theta^{\mu \nu} \rangle \equiv \eta^{\mu \nu} B = - N_c N_f \int \frac{d^4k}{(2\pi)^4} {\rm Tr} \left [ V_{gq}^{\mu \nu} S_k \right ],
 \label{eq:tmn}
\end{eqnarray}
where $S_k=i/(\slashed{k}-M+i \epsilon)$ is the Feynman quark propagator, $N_c=3$ is the number of colors, $N_f$ is the number of flavors, 
and the trace is over the Dirac 
indices. Evaluation of the trace yields 
\begin{eqnarray}
\langle \theta^{\mu \nu} \rangle  = - 4 i N_c N_f \int \frac{d^4k}{(2\pi)^4} \frac{k^\mu k^\nu - (k^2-M^2)\eta^{\mu \nu}}{k^2-M^2+i\epsilon}.
\end{eqnarray} 
Since in our low-energy model we focus on soft contributions, the perturbative contribution, 
corresponding to the above expression taken at $M=0$, is subtracted.\footnote{This amounts to using the condition 
$\int d^4k \, 1 =0$, as in the case of the dimensional regularization.}  The result is
\begin{eqnarray}
\tfrac{1}{4}\langle \theta \rangle = B=\frac{N_c  N_f}{16 \pi^2} M^2 A_0(M^2),  \label{eq:bag}
\end{eqnarray}
where  $A_0(M^2)$ is the one-point Passarino-Veltman function (cf. Appendix~\ref{sec:pave}).

Analogously, the quark condensate (for a single flavor) is
\begin{eqnarray}
\langle \bar{q} q \rangle = - N_c \int \frac{d^4k}{(2\pi)^4} {\rm Tr} \left [ V_{\sigma q} S_k \right ]=  \frac{N_c}{4\pi^2} M A_0(M^2), \label{eq:qq2}
\end{eqnarray}
Comparing~(\ref{eq:bag}) and (\ref{eq:qq2}), we notice the relation (holding at the quark-model scale)
\begin{eqnarray}
B=\frac{N_f M}{4} \langle \bar{q} q \rangle, \label{eq:rel}
\end{eqnarray}
or, equivalently,
\begin{eqnarray}
\langle \theta \rangle =M N_f \langle \bar{q} q \rangle. \label{eq:relalt}
\end{eqnarray}
The typical values of $M\sim 0.3$~GeV are compatible within uncertainties with 
the QCD sum rule estimates from charmonium decay, $B=-(224^{+35}_{-70}~{\rm MeV})^4$ \cite{Ioffe:2002ee,Ioffe:2002be} and 
$\langle \bar{q} q \rangle = (235\pm 15~\rm{MeV})^3$.
However, Eqs.~(\ref{eq:bag}, \ref{eq:qq2}) are formal in the sense that the quantities are ultraviolet divergent. 
In particular, $A_0(M^2)$ of Eq.~(\ref{eq:pave}) is quadratically divergent, hence, as is well known, 
a suitable regularization is needed to incorporate only the soft physics and make the expressions finite. The procedure disregards the 
perturbative QCD contributions, which are considered separately.
In the Pauli-Villars regularization, as introduced, e.g., in~\cite{Davidson:1994uv}, relation~(\ref{eq:rel}) holds.
In the case of the Spectral Quark Model (SQM)~\cite{RuizArriola:2003bs} (see Appendix~\ref{sec:spectral}) we find
\begin{eqnarray}
B=-\frac{N_c N_f M_V^4}{192 \pi^2}, \label{eq:sqm}
\end{eqnarray}
where $M_V\simeq 770$~MeV is the vector meson mass. Numerically,
we get $B=-(200-220~{\rm MeV})^4$ (with $N_f=3$), in agreement with the QCD sum rules estimate quoted above. 
Since in SQM~\cite{RuizArriola:2003bs} the relation
\begin{eqnarray}
f^2=\frac{N_c M_V^2}{24\pi^2} \label{eq:f2SQM}
\end{eqnarray}
holds,\footnote{It is also obtained when matching chiral
quark models to the vector-meson dominance principle, and is reasonably well satisfied 
numerically with the $M_V\simeq m_\rho$.} where $f \simeq 93~{\rm MeV}$ is the pion weak decay constant, one can also write
\begin{eqnarray}
B=-\frac{3\pi^2 N_f f^4}{N_c}. \label{eq:sqm3}
\end{eqnarray}

In SQM, the value of the quark condensate is an independent model
parameter~\cite{RuizArriola:2003bs}, unrelated to $B$.  While
phenomenologically this a welcome feature, the numerical agreement of
$B$ with QCD sum rules should be taken with care. Since $B$ contains
both quarks and gluons, it behaves as $\sim N_c^2 $, while numerically 
the scale is $\sim 1$~ GeV, whereas in the quark models we only have quarks, hence $B \sim N_c N_f $
and the scale is $\sim 0.3 {\rm GeV}$. The scale dependence of the
quark condensate confirms this feature~\cite{RuizArriola:2002wr}.

\section{Two-point functions}

The SEM correlations receive contributions from the bottom-row diagrams of Fig.~\ref{fig:tad}, from where we find
\begin{eqnarray}
&& \hspace{-7mm} \langle \theta^{\mu\nu}(q)\theta^{\alpha \beta}(-q) \rangle =  
i N_c N_f \int \frac{d^4k}{(2\pi)^4} {\rm Tr} \left [ i V_{gq}^{\mu \nu} S_{k+q} i V_{gq}^{\alpha \beta} S_{k} + i V_{ggq}^{\mu \nu,\alpha\beta} S_k  \right ]. \nonumber \\
\end{eqnarray}
The result can be decomposed into transverse spin-0 and spin-2 components, as well as a longitudinal part,
\begin{eqnarray}
\hspace{-7mm} \langle \theta^{\mu \nu}(q) \theta^{\alpha \beta}(-q)\rangle &=& \langle \theta \theta\rangle_0(q^2)  T_0^{\mu \nu \alpha \beta}+
\langle \theta \theta\rangle_2(q^2)  T_2^{\mu \nu \alpha \beta} + \langle \theta \theta\rangle_L T_L^{\mu \nu \alpha \beta}, \label{eq:thdecomp}
\label{eq:decomp}
\end{eqnarray}
where the tensors are explicitly given in Appendix~\ref{sec:ten}. The evaluation of the one-loop integrals 
yields the following expressions in terms of the Passarino-Veltman functions (\ref{eq:pave}):
\begin{eqnarray}
&& \langle \theta \theta \rangle_0(q^2) = \frac{N_c N_f}{48\pi^2} M^2   \left [ 5A_0(M^2) + 2(q^2-4M^2) B_0(M^2,q^2) \right ],\label{eq:tt02} \\
&& \langle \theta \theta \rangle_2(q^2) = \frac{N_c N_f}{480\pi^2} 
\left [ 2(46M^2-3q^2)A_0(M^2) \right . \nonumber \\
&& ~\hspace{36mm} +\left .  (q^2-4M^2)(8M^2+3 q^2) B_0(M^2,q^2) \right ], \nonumber \\
&& \hspace{5mm} \langle \theta \theta \rangle_L = -\frac{N_c N_f}{16\pi^2} M^2 A_0(M^2) = -B. \nonumber
\end{eqnarray}
The form of the longitudinal part, with the help of relation~(\ref{eq:Lprop}), clearly leads to the fulfillment of the WTI of Eq.~(\ref{eq:wti3}).
This feature is of course expected, as our calculation conforms to all the covariance requirements.

In particular, for the trace-trace correlator we find
\begin{eqnarray}\langle \theta \theta \rangle(q^2) = 
\frac{N_c N_f}{8\pi^2} M^2   (q^2-4M^2) B_0(M^2,q^2) ,\label{eq:thth} 
\end{eqnarray}

For the SEM-scalar correlator, 
following the notation of~\cite{Donoghue:1991qv}, we have the general decomposition into the transverse and longitudinal parts,
\begin{eqnarray}
\langle \theta^{\mu\nu}(q)\bar{q}q(-q) \rangle
  =  t^{\mu \nu} q^2 \phi_1(q^2)+ \eta^{\mu \nu} \phi_0 (q^2). \label{eq:sca}
\end{eqnarray}
At the one-quark-loop level
\begin{eqnarray}
\hspace{-7mm} \langle \theta^{\mu\nu}(q)\bar{q}q(-q) \rangle =  
i N_c N_f \int \frac{d^4k}{(2\pi)^4} {\rm Tr} \left [ i V_{gq}^{\mu \nu} S_{k+q} i V_{\sigma q} S_{k} + i V_{g\sigma q}^{\mu \nu} S_k  \right ],
\end{eqnarray}
yielding
\begin{eqnarray}
&&\hspace{-7mm}q^2 \phi_1=\frac{N_c M}{24\pi^2} \left [(q^2-4M^2)B_0(M^2,q^2) + 4 A_0(M^2)\right ], \nonumber \\
&&\hspace{-7mm}\phi_0=-\frac{N_c M}{4\pi^2}  A_0(M^2) = -\langle \bar{q} q \rangle. \label{eq:scal}
\end{eqnarray}
We note that since $\phi_0=-\langle \bar{q} q \rangle$, the WTI of Eq.~(\ref{eq:wtiqq}) is immediately satisfied.

The structure of the one-loop correlators
(\ref{eq:tt02}, \ref{eq:scal}) incorporates absorptive parts, contained
in the $B_0$ function (cf. Eq.~(\ref{eq:disc})) and appearing in the
transverse pieces only.  The remaining dispersive pieces are polynomial
in $q^2$. Note that the perturbative renormalization may redefine the
polynomial terms but not the absorptive pieces, which are in that sense
unique, since they represent a branch cut discontinuity at $q^2= 4
M^2$.  In the spectral regularization, the polynomial pieces $q^2$ and $q^4$ with arbitrary coefficients emerge only in $\langle \theta \theta \rangle_2$, as is apparent 
from Eqs.~(\ref{eq:momexpl0}, \ref{eq:momexpl}), and not in the other considered correlators.

The low-energy model considered here is by construction valid at soft momenta, which also concerns the value of $q^2$.
Yet, we notice that with the relation~(\ref{eq:asrel}) holding in  SQM,
we readily find that at $q^2\to \infty$ the transverse part 
of  $\langle \theta^{\mu\nu} \bar{q}q \rangle(q^2)$
tends to 
\begin{eqnarray}
q^2\phi_1(q^2) \to \frac{N_c M}{4\pi^2} A_0(M^2) = \langle \bar{q} q \rangle.
\end{eqnarray}
Incidentally, this limit agrees with Eqs.~(65, 66) of~\cite{Donoghue:1991qv}.

Explicit expression in SQM for all the considered correlators are listed in Appendix~\ref{sec:corrSQM}.

\section{Trace identity for SEM in the quark model}

In our quark model valid at the quark model scale 
no anomalies are considered, $\mathcal{A}=0$. Moreover,  $\langle \theta^\mu_\mu\rangle =M N_f \langle {\bar q}q \rangle$ in Eq. (\ref{eq:tr0}).
Then Eq.~(\ref{eq:trid}) simplifies into
\begin{eqnarray}
\eta_{\mu\nu} \langle \theta^{\mu\nu}(q) \theta^{\alpha\beta}(-q) \rangle =  M N_f \langle [\bar q q](q) \theta^{\alpha\beta}(-q)\rangle
+ 2\, \langle \theta^{\alpha\beta} \rangle. \label{eq:tridq}
\end{eqnarray}
This relation, binding non-trivially the SEM-SEM correlator and the SEM-scalar correlator, is explicitly satisfied by 
our quark-model expressions (\ref{eq:thdecomp}, \ref{eq:tt02}, \ref{eq:sca}, \ref{eq:scal}), where formulas (\ref{eq:tcont}) for the contracted tensors are used.

\section{Low-energy theorems vs current conservation}

In QCD, the trace operator $\theta=\theta^\mu_\mu$ plays an important role. 
It involves the anomaly and is related to the divergence of the dilatation current $D(x)$ at the quantum level,
\begin{eqnarray}
\partial^\mu D_\mu = \theta   = 
\frac{\beta(\alpha)}{2\alpha}
G^{\mu\nu a} G_{\mu\nu}^a + \sum_q m_q \left[ 1 + \gamma_m (\alpha) \right] \bar q q, 
\label{eq:anom}
\end{eqnarray} 
where  $\beta(\alpha) = \mu^2 d \alpha / d\mu^2 $ is the QCD beta
function at an energy scale $\mu$, $\alpha=g^2/(4\pi)$ is the running coupling constant, $G$ is the gluon field, and
$\gamma_m(\alpha) = d \log m_q / d \log \mu^2 $ denotes the anomalous
dimension of the current quark mass $m_q$, with $q$ summed over the flavors. The trace anomaly $\theta^\mu_\mu$ is renormalization-scheme and scale independent, being free of the UV divergences.

With certain assumptions concerning regularization and renormalization, the following low-energy theorems (LET) or sum rules
have been obtained~\cite{Novikov:1981xi}:
\begin{eqnarray}
i \int d^4 x \langle T \theta(x) O(0)\rangle = - d_O \langle O \rangle,  \label{eq:sumr}
\end{eqnarray}
where $O$ is an arbitrary operator with an (engineering) dimension $d_O$. The case of $O=\bar q q$ from the point of view of LET 
was discussed in detail in Sec.~VIII of~\cite{Donoghue:1991qv}. 

In particular, for the SEM-scalar correlator Eq.~(\ref{eq:sumr}) would mean that (in the momentum space at $q=0$)
\begin{eqnarray}
\langle \theta(0)\bar{q}q(0) \rangle=-3 \langle \bar{q}q \rangle, \hspace{3cm} {\rm (LET)} \label{eq:LETqq}
\end{eqnarray}
with $d_{\bar q q}=3$.
We note, however, that this formula is incompatible with WTI. Indeed,  by contracting Eq.~(\ref{eq:sca}) with $q_\mu$
and comparing to WTI of Eq.~(\ref{eq:wtiqq}), one readily finds that $\phi_0=-\langle \bar{q}q \rangle$. The result follows 
from the conservation of SEM, hence is based on symmetry preservation and, thus, general.
Taking the trace of Eq.~(\ref{eq:sca}) yields 
\begin{eqnarray}
\langle \theta(q)\bar{q}q(-q) \rangle =  3 q^2 \phi_1(q^2)+ 4 \phi_0 (q^2).
\end{eqnarray}
Since $\phi_1$ cannot have a massless pole, this means that 
\begin{eqnarray}
\langle \theta(0)\bar{q}q(0) \rangle=-4 \langle \bar{q}q \rangle. \hspace{3cm} {\rm (WTI)} \label{eq:scalf}
\end{eqnarray}
The LET result of Eq.~(\ref{eq:LETqq}) demands $\phi_0=-\tfrac{3}{4}\langle \bar{q}q \rangle$~\cite{Donoghue:1991qv}, in 
tension with WTI.

Similarly, according to Eq.~(\ref{eq:sumr}), LET for the trace-trace correlator reads
\begin{eqnarray}
\langle \theta(0)\theta(0) \rangle=-4 \langle \theta \rangle, \hspace{3cm} {\rm (LET)} \label{eq:LETSEM}
\end{eqnarray}
with $d_\theta=4$.
On the other hand, WTI of Eq.~(\ref{eq:wti4}) implies immediately
\begin{eqnarray}
q_\mu \langle \theta(q)\theta^{\mu\nu}(-q) \rangle=-2 q^\nu B.  \label{eq:genSEM}
\end{eqnarray}
The generic structure of the correlator is therefore 
\begin{eqnarray}
\langle \theta(q)\theta^{\mu\nu}(-q) \rangle=-2 g^{\mu\nu} B + (q^\mu q^\nu-g^{\mu \nu}q^2)C(q^2).  \label{eq:genSEM3}
\end{eqnarray}
With no massless pole in $C(q^2)$, in the $q^2\to 0$ limit we find the relation
\begin{eqnarray}
\lim_{q^2\to 0}\langle \theta(q)\theta(-q) \rangle=-8  B = -2\langle \theta \rangle. \hspace{3cm} {\rm (WTI)} \label{eq:finSEM4}
\end{eqnarray}

We stress that the relations (\ref{eq:scalf}) and (\ref{eq:finSEM4}) follow directly from the WTI (current conservation), i.e., from the Lorentz covariance, 
supplemented with the physical requirement of the absence of zero-mass poles, hence are general, model independent results.

The explicit results obtained in our quark model naturally comply to Eqs.~(\ref{eq:scalf}) and (\ref{eq:finSEM4}). We note that they are also 
compatible with the trace identity (\ref{eq:tridq}). 
Formula~(\ref{eq:finSEM4}), with the proper factor of $-2$ on the right-most side, compared to $-4$ in Eq.~(\ref{eq:LETSEM}),
agrees with the sum rule obtained by Brown and Zee~\cite{Brown-Zee,Donoghue:2017vvl}.

For the $\langle \theta \theta \rangle$ case in QCD, the intriguing conflict between the LET results and the Lorentz symmetry constraints has been explained in~\cite{Donoghue:2017vvl}.

\section{Connection to induced gravity} 

The Adler-Zee formula~\cite{Adler:1980pg,Zee:1980sj} for the induced~\cite{Adler:1980pg,Zee:1980sj,Adler:1982ri,Zee-2} Newton's gravitational 
constant reads (cf. Eq.~(15) in~\cite{Donoghue:2017vvl})
\begin{eqnarray}
\frac{1}{16\pi G_{\rm ind} }=  \left . \frac{1}{12}\frac{d}{dq^2} \langle \theta(q)\theta(-q) \rangle \right |_{q^2=0}. \label{eq:AZ}
\end{eqnarray}
In SQM (see also~\cite{Megias:2004uj,Megias:2005fj}), with the help of  Eqs.~(\ref{eq:lowq}) and (\ref{eq:tcont}), we find
\begin{eqnarray}
\frac{1}{16\pi G_{\rm ind} }=-\frac{\langle \theta \rangle}{12M_V^2}= \frac{N_c N_f M_V^2 }{576 \pi ^2} \simeq 0.001~{\rm GeV}^2. \label{eq:Gind}
\end{eqnarray}
Only the scalar component $\langle \theta \theta \rangle_0(q)$ contributes to Eq.~(\ref{eq:AZ}).
With the relation (\ref{eq:f2SQM}), Eq.~(\ref{eq:Gind}) can also be written as 
\begin{eqnarray}
\frac{1}{16\pi G_{\rm ind} }= \frac{N_f f^2 }{24}. \label{eq:Gindf}
\end{eqnarray}

The induced cosmological 
constant\footnote{We use the convention ${\cal L}_{\rm grav}=\frac{1}{16\pi G}\int d^4 x(R-2\Lambda)$.} is equal to (cf.~Eq.~(5.35) of~\cite{Adler:1982ri})
\begin{eqnarray}
\Lambda_{\rm ind}=-2\pi G_{\rm ind}  \langle \theta \rangle= -8\pi G_{\rm ind} B. 
\end{eqnarray}
In SQM, according to Eq.~(\ref{eq:sqm}) and (\ref{eq:Gind}), we get 
\begin{eqnarray}
\Lambda_{\rm ind}=\frac{3M_V^2}{2}=\frac{36\pi^2 f^2}{N_c}\simeq 1~{\rm GeV}^2.
\end{eqnarray}
The huge mismatch compared to the observed value of $\Lambda \simeq 10^{-84}\,{\rm GeV}^2$ is a manifestation of  the well known 
cosmological constant problem~\cite{Weinberg:1988cp,Martin:2012bt}, experienced in Quantum Field Theory.

Similarly, the numerical value of $G_{\rm ind}$ from Eq.~(\ref{eq:Gind}) is 
\begin{eqnarray}
G_{\rm ind} \simeq 20~{\rm GeV}^{-2},
\end{eqnarray}
compared to the experimental value of the Newton gravitational constant, $G\simeq 6.7\times 10^{-39}~{\rm GeV}^{-2}$.

\section{Dispersion relations and temporal correlations}

Temporal correlations of two operators, $O_1$ and $O_2$ (such as used in lattice QCD studies~\cite{Meyer:2009jp,Taniguchi:2017ibr}), are defined as 
\begin{eqnarray}
\hspace{-7mm} \langle O_1 O_2 \rangle (t)&\equiv& i \int d^3x \langle 0|TO_1(\vec{x},t)O_2(0)|0\rangle  = \int \frac{dq_0}{2\pi} e^{i q_0 t}  \langle O_1 O_2 \rangle(q_0^2). \label{eq:temp}
\end{eqnarray}
In our case $O_i$ are the SEM tensors or $\bar q q$. These correlators, being analytic functions, satisfy suitably subtracted dispersion relations. 
However, it is clear from Eq.~(\ref{eq:temp}) that the polynomial pieces in $q_0$ generate the Dirac $\delta(t)$ function and its derivatives, hence 
they do not show up at non-zero $t$.
Therefore, in the analysis of the temporal SEM correlations we only need the discontinuities of Eqs.~(\ref{eq:tt02}, \ref{eq:scal}) residing in the 
Passarino-Veltman $B_0$ functions, which have the explicit form
\begin{eqnarray}
&& {\rm disc} \,\Pi_0(s) = M N_f s \,{\rm disc} \, \phi_1(s) = \nonumber \\
&& \hspace{22mm} \frac{2\pi i N_c N_f}{24\pi^2}   \theta(s-4M^2)\sqrt{1-\frac{4M^2}{s}} M^2(s-4M^2),  \label{eq:ttdiscoM} \\ 
&& {\rm disc} \,\Pi_2(s)= \frac{2\pi i N_c N_f}{480\pi^2}  \theta(s-4M^2)\sqrt{1-\frac{4M^2}{s}} (s-4M^2)(3s+8M^2),  \nonumber 
\end{eqnarray}
where $s=q^2=q_0^2$.
Essentially, the model provides the origin of the $q \bar{q}$ production cut at $s=4M^2$ and a specific dependence on $M$ in the
multiplying polynomials.

Taking the Fourier transform of (\ref{eq:ttdiscoM}) yields, at large Euclidean times $\tau = i t$, the behavior 
\begin{eqnarray}
\Pi_{0,2} (\tau) &\sim& \frac{N_c N_f M^3 e^{-2M \tau }}{12 \pi^2 \tau^2 },   \label{eq:tempoasM} \\
\phi_1(\tau) &\sim& \frac{N_c e^{-2M \tau }}{48 \pi^2 \tau^2 }. \nonumber
\end{eqnarray}
A similar asymptotic form for $\Pi_{0}(\tau)$, $\Pi_{2}(\tau)$, and $\phi_1(\tau)$ reflects the same limit of the corresponding discontinuities in (\ref{eq:ttdiscoM}) 
near the $\bar q q$ production threshold $s=4M^2$. The slowly-varying function, above of the form $\sim 1/\tau^2$, is altered with regularization. 
In particular, in SQM we obtain
\begin{eqnarray}
\Pi_{0,2} (\tau) &\sim&\frac{N_c N_f M_V^4 e^{-M_V \tau }}{192 \pi^3 \tau },   \label{0eq:tempoas} \\
\phi_1(\tau) &\sim&-\frac{\langle \bar q q \rangle e^{-M_S \tau }}{4 \pi M_S^2 \tau }. \nonumber
\end{eqnarray}

It would be a bit naive to expect that the nuances of the slowly varying functions in the temporal correlators could be sorted out on the lattice, 
where one expects meson dominance in channels with the corresponding quantum numbers. 
Nevertheless, the derived expressions can be used to estimate the magnitude of the non-resonant (constituent) quark continuum to these correlators.

\section{Conclusions}

In  this paper
we have considered, using standard procedures of the path integral formulation of field theory
in curved space-time, the Ward-Takashi and trace identities for the two-point
correlators (Green's functions) of the stress-energy-momentum tensor
for a general non-conformal case.  The identities were then
illustrated on a  relevant example of the free massive Dirac fermion theory,
treated at the one-loop level. Physically, the model corresponds to
chiral (constituent) quarks, used in numerous phenomenological applications in the soft
non-perturbative regime of QCD. We have discussed the constraints
following from WTIs  in reference to the low-energy sum rules involving the trace of the stress-energy-momentum tensor at zero momentum, 
underlying the key role of these symmetry constraints. 
We have also obtained the temporal representation
of the two-point correlators in the quark model, which provides a  possibly useful
benchmark for the non-perturbative quark continuum contribution in future lattice QCD
studies. 

\bigskip

ERA was supported by Grant No. PID2023-147072NB-I00 funded by MICIU/AEI/10.13039/501100011033
and by the ERDF/EU, and by Grant No. FQM-225 funded by the Junta de Andalucía.

\appendix

\section{Tensors \label{sec:ten}}

We introduce the basic tensors appearing in our analysis, where $q$ is the graviton momentum. Let
\begin{eqnarray}
t^{\mu \nu}=\eta^{\mu \nu}-q^{\mu \nu}, \;\; q^{\mu \nu}=\frac{q^\mu q^\nu}{q^2}, \label{eq:t}
\end{eqnarray}
with $t^{\mu \nu}q_\nu=0$. The spin 0 and 2 combinations form the basis for the transverse parts of SEM, 
\begin{eqnarray}
&& \hspace{-7mm} T_0^{\mu \nu \alpha \beta}= \tfrac{1}{3} t^{\mu \nu}t^{\alpha \beta}, \;\; T_2^{\mu \nu \alpha \beta}= 
\tfrac{1}{2} t^{\mu \alpha}t^{\nu \beta}+ \tfrac{1}{2} t^{\mu \beta}t^{\nu \alpha}-\tfrac{1}{3} t^{\mu \nu}t^{\alpha \beta}. \nonumber \\ \label{eq:t02}
\end{eqnarray}
In addition, we need the tensor appearing in the longitudinal part of SEM, 
\begin{eqnarray}
T_L^{\mu \nu \alpha \beta}&=&q^{\alpha  \beta} q^{\mu  \nu }+\eta ^{\mu  \nu } q^{\alpha  \beta }+\eta^{\alpha  \beta } q^{\mu  \nu }- \\
&& \eta ^{\beta  \nu } q^{\alpha  \mu }-\eta ^{\alpha  \nu }q^{\beta  \mu }-\eta ^{\beta  \mu } q^{\alpha  \nu }-\eta ^{\alpha  \mu } q^{\beta  \nu }, \nonumber
\end{eqnarray}
with  the property 
\begin{eqnarray}
q_\nu T_L^{\mu \nu \alpha \beta} = q^\mu \eta^{\alpha \beta} - q^\alpha \eta^{\mu \beta} - q^\beta \eta^{\mu \alpha}. \label{eq:Lprop}
\end{eqnarray}
The transverse tensors exhibit the projective property
\begin{eqnarray}
&& T_0^{\mu \nu \alpha \beta} {T_{0 \; \alpha \beta}}^{\sigma \tau}=T_0^{\mu \nu \sigma \tau}, \nonumber \\
&& T_2^{\mu \nu \alpha \beta} {T_{2 \; \alpha \beta}}^{\sigma \tau}=T_2^{\mu \nu \sigma \tau}, \nonumber \\
&& T_0^{\mu \nu \alpha \beta} {T_{2 \; \alpha \beta}}^{\sigma \tau}=0. \label{eq:proj}
\end{eqnarray}
The tensors contracted with $\eta_{\mu \nu}$ become
\begin{eqnarray}
&& \eta_{\mu \nu} T_0^{\mu \nu \alpha \beta} = t^{\alpha \beta}, \nonumber \\
&& \eta_{\mu \nu} T_2^{\mu \nu \alpha \beta} =0, \nonumber \\
&& \eta_{\mu \nu} T_L^{\mu \nu \alpha \beta} = \eta^{\alpha \beta}+q^{\alpha \beta}, \label{eq:tcont} \\
&& \eta_{\mu \nu} \eta_{\alpha \beta} T_0^{\mu \nu \alpha \beta} =3, \;\; \; \eta_{\mu \nu} \eta_{\alpha \beta} T_L^{\mu \nu \alpha \beta} =5. \nonumber
\end{eqnarray}

\section{Passarino-Veltman functions \label{sec:pave}}

The adopted convention for the Passarino-Veltman one-loop functions is
\begin{eqnarray}
&& \hspace{-7mm}  i \pi^2 A_0(M^2) =\int\!\! d^4k \frac{1}{k^2-M^2+i \epsilon}, \label{eq:pave} \\
&& \hspace{-7mm}  i \pi^2 B_0(M^2, q^2) =\!\!\int\!\! d^4k \frac{1}{[(k+q)^2\!-\!M^2\!+\!i\epsilon][k^2\!-\!M^2\!+\!i\epsilon]}, \nonumber
\end{eqnarray} 
or, with the Euclidean momenta upon the Wick rotation $K_0=-i k_0$, 
\begin{eqnarray}
&&  \pi^2 A_0(M^2) =- \int\!\! d^4K \frac{1}{K^2+M^2}, \label{eq:paveE} \\
&&  \pi^2 B_0(M^2, q^2=-Q^2) =\!\!\int\!\! d^4K \frac{1}{[(K+Q)^2\!+\!M^2][K^2\!+\!M^2]}. \nonumber
\end{eqnarray} 
These functions are, correspondingly, quadratically and logarithmically divergent. With the Feynman parametrization of the denominators one formally finds 
\begin{eqnarray}
B_0(M^2, q^2) = - \int_0^1 dx \log \left ( \frac{M^2 -x(1-x)q^2}{\lambda^2} \right ),~~~~ \label{eq:B0log}
\end{eqnarray} 
where $\lambda$ is a constant. The discontinuity of $B_0$ is well defined (finite) and equals to
\begin{eqnarray}
\hspace{-10mm} {\rm disc}\,B_0(M^2, q^2) = 2 i\, {\rm Im}\,B_0(M^2, q^2) =2 \pi i  \, \theta(q^2 - 4M^2) \sqrt{1-\frac{4M^2}{q^2}}. \label{eq:disc}
\end{eqnarray}

\section{Spectral regularization \label{sec:spectral}}

\begin{figure}[tb]
\begin{center}
\includegraphics[angle=0,width=0.42 \textwidth]{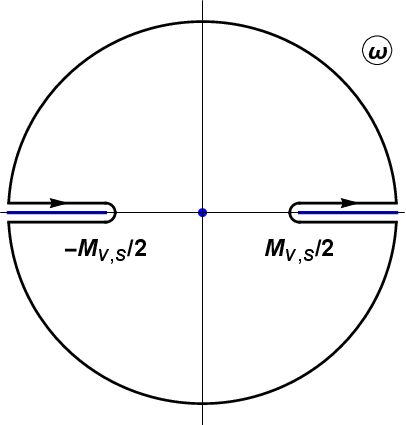} 
\end{center}
\vspace{-5mm}
\caption{Contour $C$ of the integration in the complex quark-mass plane for the spectral regularization method. 
The cuts, starting at $\pm \tfrac{1}{2}M_{V,S}$,  and the pole at $\omega=0$, present in the case of $\rho_V(\omega)$, are indicated. 
The circular part of the contour $C$ tends to infinity. \label{fig:cont}} 
\end{figure}

The details and a physical justification of the method defining SQM are provided in~\cite{RuizArriola:2003bs}. The spectral function for quarks of a (complex) mass $\omega$ 
(it corresponds to the mass $M$ in the body of the text) is 
\begin{eqnarray}
\rho(\omega) &=& \rho_V(\omega)+\rho_S(\omega), \nonumber \\
\rho_V(\omega) &=&\frac{1}{2\pi i} \frac{1}{\omega}\frac{1}{(1-4\omega^2/M_V^2)^{5/2}}, \nonumber \\
\rho_S(\omega) &=&\frac{1}{2\pi i} \frac{-48\pi^2 M_S \langle \bar{q}q\rangle }{N_c (1-4\omega^2/M_S^2)^{5/2}}, \label{eq:rho}
\end{eqnarray} 
where $M_V$ and $M_S$ are model parameters (vector and scalar masses, fixed with certain physical observables~\cite{RuizArriola:2003bs}),
accompanied with the complex-plane integration contour shown in Fig.~\ref{fig:cont}. The $V$ and $S$ parts give, correspondingly, 
the even and odd moments of the Passarino-Veltman functions in $\omega$, as listed below. 

To regularize the quark-model formulas for the one- and two-body functions appearing in the body of this paper, several lowest order
spectral moments of the Passarino-Veltman functions~(\ref{eq:paveE}) are needed, denoted as 
\begin{eqnarray}
&& \langle  w^n A_0 (w^2) \rangle = \int_C d\omega \rho(\omega) \omega^n A_0(\omega^2), \nonumber  \\
&& \langle  w^n B_0 (w^2,q^2) \rangle = \int_C d\omega \rho(\omega) \omega^n B_0(\omega^2,q^2). \label{eq:momdef}
\end{eqnarray}
For $A_0$, we obtain them from the Cauchy theorem as a sum of residues at $\omega=\pm i \tfrac{1}{2}M_{V,S}$ and also, for the case of the $n=0$ moment, at $\omega=0$, 
cf.~the $1/\omega$ pole in $\rho_V$ in  Eq.~(\ref{eq:rho}). Explicitly,
\begin{eqnarray}
\langle  w A_0 (w^2)\rangle =\frac{4\pi^2 \langle \bar{q} q \rangle}{N_c}, \;\;\; \langle  w^2 A_0 (w^2)\rangle = -\frac{M_V^4}{12}. \label{eq:momA}
\end{eqnarray}
For the $n=0$ moment of $A_0$, the residue at $\omega=0$ yields an extra (perturbative) contribution which is quadratically divergent,
\begin{eqnarray}
-\frac{1}{\pi^2}\int d^4K \frac{1}{K^2}. \label{eq:tad}
\end{eqnarray} 
Upon renormalization, it brings an arbitrary constant $\mu_A$ of the dimension of the energy squared, hence we can write
\begin{eqnarray}
\langle A_0 (w^2) \rangle = \frac{M_V^2}{6} + \mu_A^2, \label{eq:momexpl0}
\end{eqnarray}
where the first term is the non-perturbative contribution.

Similarly, for the even moments of $B_0$ we find
\begin{eqnarray}
\langle B_0 (w^2,q^2) \rangle &=&  \frac{2}{3}\frac{M_V^2}{M_V^2-q^2}-\log \left( \frac{M_V^2-q^2}{\mu_B^2} \right ),  \label{eq:momexpl} \\
\langle w^2 B_0 (w^2,q^2) \rangle &=& \frac{1}{6}\frac{M_V^4}{M_V^2-q^2}, \nonumber \\
\langle w^4 B_0 (w^2,q^2) \rangle &=& 
\frac{1}{48} M_V^4 \left[\frac{2 M_V^2}{M_V^2-q^2}-\frac{3M_V}{q} \log \left(\frac{M_V+q}{M_V-q}\right) \right], \nonumber
\end{eqnarray}
where $\mu_B$ is an arbitrary energy scale originating from the renormalization of the perturbative contribution in $\langle B_0 (w^2,q^2) \rangle$.
For the odd moments we have
\begin{eqnarray}
\langle w B_0 (w^2,q^2) \rangle &=& -\frac{8\pi^2 \langle \bar{q}q \rangle}{N_c(M_S^2-q^2)}, \label{eq:momexpl2} \\
\langle w^3 B_0 (w^2,q^2) \rangle &=& 
-\frac{\pi^2 \langle \bar{q}q \rangle}{N_c} \left[\frac{2 M_S^2}{M_S^2-q^2}-\frac{3M_S}{q} \log \left(\frac{M_S+q}{M_S-q}\right) \right]. \nonumber
\end{eqnarray}

Various low- and high-$q^2$ limits of the above formulas can be readily obtained. In particular, at $q^2\to\infty$, one finds
\begin{eqnarray}
q^2 \langle \omega B_0(\omega^2,q^2)\rangle =2 \langle \omega A_0(\omega^2)\rangle+{\cal O}(1/q^2). \label{eq:asrel}
\end{eqnarray} 

\section{Correlators in the Spectral Quark Model \label{sec:corrSQM}}

The explicit formulas for the SEM-SEM correlators of Eq.~(\ref{eq:tt02}) in SQM, with expressions for the spectral moments from Appendix~\ref{sec:spectral} inserted, 
acquire the simple forms
\begin{eqnarray}
&& \langle \theta \theta \rangle_0(q^2) = 
\frac{\langle \theta \rangle}{4} \left [ -\frac{2M_V}{q} \log \left ( \frac{M_V+q}{M_V-q} \right ) +3\right ] \nonumber \\
&& \langle \theta \theta \rangle_2(q^2) = \frac{\langle \theta \rangle}{10} 
\left [ 3 \frac{q^4}{M_V^4} \log \left ( \frac{M_V^2-q^2}{\mu_B^2} \right ) \right . \nonumber \\
&& \hspace{17mm} - \left . \frac{2M_V}{q}\log \left ( \frac{M_V+q}{M_V-q} \right ) + 9 + q^2 \frac{\nu_A^2}{M_V^4} \right ], \label{eq:ttsp} \\
&& \langle \theta \theta \rangle_L(q^2) = \frac{\langle \theta \rangle}{4} =-B, \nonumber
\end{eqnarray}
where Eq.~(\ref{eq:sqm}) has been used. 
Here $\nu_A^2=6\mu_A^2+3M_V^2$ (see Appendix~\ref{sec:spectral})  and $\mu_B$ are renormalization constants for the perturbative contributions.
Note that the terms with the pole at $q^2=M_V^2$, present in Eq.~(\ref{eq:momexpl}), cancel out, while the branch cuts appear. The 
perturbative contributions show up only in $\langle \theta \theta \rangle_0$, with the renormalization constants giving rise to the 
arbitrary polynomial pieces $q^2$ and $q^4$.

Analogously to the above expressions, for the SEM-scalar correlator we get
\begin{eqnarray}
&& q^2 \phi_1(q^2)=\langle \bar{q} q \rangle \left [ 1-\frac{M_S}{2q} \log \left ( \frac{M_S+q}{M_S-q} \right )\right ], \nonumber \\
&& \phi_0(q^2)=-\langle \bar{q} q \rangle. \label{eq:phi1sqm}
\end{eqnarray}

The low-momentum expansion of the above functions yields
\begin{eqnarray}
&& \langle \theta \theta \rangle_0(q^2) = \langle \theta \rangle \left [-\frac{1}{4}-\frac{q^2}{3M_V^2} +{\cal O}(q^4/M_V^4)\right ], \nonumber \\
&& \langle \theta \theta \rangle_2(q^2) = \langle \theta \rangle \left [ \frac{1}{2} +\frac{q^2(18\mu_A^2+5M_V^2)}{30M_V^2} +{\cal O}(q^4/M_V^4)\right ] ,\label{eq:lowq} \\
&& \phi_1(q^2)= - \langle \bar q q \rangle \left [ \frac{1}{3M_S^2}+ \frac{q^2}{5M_S^4}+{\cal O}(q^4/M_S^4) \right ], \nonumber
\end{eqnarray} 
while at asymptotically high $Q^2=-q^2$ we find 
\begin{eqnarray}
&& \langle \theta \theta \rangle_0(q^2) =  \frac{3}{4}\langle \theta \rangle +{\cal O}(M_V/|q|), \nonumber \\
&& \langle \theta \theta \rangle_2(q^2) =   \langle \theta \rangle 
\left[ \frac{3q^4}{10M_V^4} \log (-q^2/\mu_B^2)+ \frac{3\mu_A^2}{5M_V^4} q^2+\frac{3}{4}+{\cal O}(M_V/|q|)\right ], \nonumber \\
&&  q^2 \phi_1(Q^2)=\langle \bar{q} q \rangle (1+{\cal O}(M_S/|q|). \label{eq:highq}
\end{eqnarray} 

The leading behavior of the spin-2 (traceless) term can  be rewritten as 
\begin{eqnarray}
\langle \theta \theta \rangle_2(q^2)=- \frac{N_c N_f}{160 \pi^2} \, q^4 \log(-q^2/\mu_B^2) + {\cal O}(q^2/M_V^2),
\end{eqnarray}
which agrees with the leading quark term in 
OPE (see, e.g., Eq.~(4.1) in~\cite{Zoller:2012qv}).\footnote{Note our convention for the $T_2$ tensor in Eq.~(\ref{eq:t02}).}
This is not incidental, as we incorporate the perturbative contribution via the massless pole in the quark spectral function~(\ref{eq:rho}). 
In the spin-0 (traceful) or the longitudinal correlators, such  terms do 
not appear in a non-interacting theory  due to the conformal symmetry in the massless quark limit ($M_V, M_S \to 0$).

We also note that the large dynamical scales $M_V$ and $M_S$ allow for 
terms in the large-momentum expansion which are absent in the standard OPE.


\bibliography{refs-barpi.bib}

\end{document}